# Gate-activated photoresponse in a graphene p-n junction

Max C. Lemme[1*], Frank H.L. Koppens[1,4*], Abram L. Falk[1], Mark S. Rudner[1], Hongkun Park[1,2], Leonid S. Levitov[3], Charles M. Marcus[1]

[1]*Department of Physics, Harvard University, Cambridge, MA 02138, USA*
[2]*Department of Chemistry and Chemical Biology, Harvard University, Cambridge, MA 02138, USA*
[3]*Department of Physics, Massachusetts Institute of Technology, Cambridge, MA 02139, USA*
[4] *ICFO-The Institute of Photonic Sciences, 08860 Castelldefels (Barcelona), Spain*
*These authors contributed equally.

ABSTRACT

We study photodetection in graphene near a local electrostatic gate, which enables active control of the potential landscape and carrier polarity. We find that a strong photoresponse only appears when and where a p-n junction is formed, allowing on-off control of photodetection. Photocurrents generated near p-n junctions do not require biasing and can be realized using submicron gates. Locally modulated photoresponse enables a new range of applications for graphene-based photodetectors including, for example, pixilated infrared imaging with control of response on subwavelength dimensions.

MANUSCRIPT TEXT

Graphene is a promising photonic material[1] whose gapless band structure allows electron-hole pairs to be generated over a broad range of wavelengths, from UV, visible[2], and telecommunication bands, to IR and THz frequencies[3]. Previous studies of photocurrents in graphene have demonstrated photoresponse near metallic contacts[4-7], at the interface between single-layer and bilayer regions[8], or at the edge of chemically doped regions[10]. Photocurrents generated near metal contacts were attributed to electric fields in the graphene that arise from band bending near the contacts[5-7], and could be modulated by sweeping a global back-gate voltage with the potential of the contacts fixed. In these studies, photocurrent away from contacts and interfaces was typically very weak. In contrast, the present study concerns devices with top gates, separated from otherwise homogeneous graphene by an insulator, $Al_2O_3$, deposited by atomic layer deposition (ALD). When the top gate inverts the carrier type under the gate, a *p-n* junction is formed at the gate edges, and a highly localized photocurrent is observed using a focussed scanning laser. A density difference induced by the top gate that does not create a *p-n* junction does not create local photosensitivity.

---

* These authors contributed equally to this work.

Comparing experimental results to theory suggests that the photocurrent generated at the *p-n* interface results from a combination of direct photogeneration of electron-hole pairs in a potential gradient, and a photothermoelectric effect in which electric fields result from optically induced temperature gradients[8,11]. Both effects are strongly enhanced at *p-n* interfaces: The enhancement of direct photocurrent results from its scaling inversely with local conductivity, while the thermoelectric contribution is enhanced by the strong spatial dependence of the Seebeck coefficient near the *p-n* interface. As neither mechanism is wavelength selective, the overall effect should provide broadband photosensitivity. We further anticipate that the ability to activate local photosensitive regions using gate voltages will provide pixel-controlled bolometers for imaging or spectroscopy with broadband sensitivity and subwavelength spatial resolution.

A typical device layout and micrograph are shown in Fig. 1. Graphene was deposited onto ~300 nm of silicon dioxide on degenerately doped silicon by mechanical exfoliation, similar to the method described by Novoselov et al.[12]. Contacts (titanium/gold) to graphene were defined by conventional electron beam lithography, and a functionalization layer based on $NO_2$ was deposited by atomic layer deposition (ALD), followed by *in situ* ALD of 20 nm of aluminum oxide ($Al_2O_3$) using a trimethylaluminum precursor [13,14]. Finally, the gate electrodes were defined by electron beam lithography and deposited by Ti / Au (5 nm / 40 nm) thermal evaporation.

Devices were characterized initially in vacuum, in a standard field-effect transistor (FET) configuration with a source-drain bias of $V_D = 1$ mV as a function of top and back gate (substrate) voltages. A two-dimensional plot of drain current $I_D$ as a function of top gate voltage, $V_T$, and back gate voltage, $V_{BG}$, for the device in Fig. 1b is shown in Fig. 1c, with white lines indicating charge neutrality points under and outside of the top-gated region. The four regions defined by these lines are denoted *p-n-p*, *n-n'-n*, *p-p'-p*, and *n-p-n*, with the middle letter indicating the region under the top gate.

After electrical testing, the devices were wire bonded to chip carriers and placed in a chip socket for high resolution scanning photocurrent measurements. These were carried out using a custom-built confocal scanning microscope. The excitation source (Koheras SuperK supercontinuum laser) was coupled to an acousto-optic tunable filter, enabling the excitation wavelength to be tuned through the visible spectrum. The beam was directed into a 100x objective using a scanning mirror. The objective lens of the microscope (100x, 0.8 numerical aperture) focused the beam to a diffraction-limited spot on the device of about 500 nm diameter. All measurements were taken at room temperature in ambient atmosphere. This measurement condition is different from the initial testing and the background doping was changed from slight *p*-type to *n*-type. Figure 1d shows the photocurrent response to the scanned laser for zero applied source-drain bias and gate voltages, $V_D = V_T = V_{BG} = 0$. In these scans, the laser spot size was ~ 0.5 μm, with wavelength λ = 600 nm and power P = 40 μW. Red and blue regions in the figure close to the source and drain contacts and on either side of the gate electrode represent distinct electron and hole photocurrents.

Photocurrent as a function of top gate voltage and position (along the vertical cut in Fig. 1d) is shown in Fig. 2a. A strong photoresponse on the two sides of the top gate appear for $V_T < V_{Dirac} = 0.9$ V, corresponding to the Dirac point under the top gate. For $V_T < V_{Dirac}$ holes are the majority carrier type under the top gate, while for $V_T > V_{Dirac}$ electrons dominate. Taking into account the slight n-type background doping of the graphene flake this indicates that the photoresponse near the gate is strong for an *n-p-n* configuration, and absent for an *n-n'-n* configuration. We note that no appreciable photocurrent was observed in the range $V_T = 1.7$ V to 10 V (not shown in the figure). Comparable gate-dependent localized photoresponse was observed in several devices over a range of excitation wavelengths from 480 nm to 750 nm.

In recent experiments[5-7], photocurrents induced by laser illumination near metallic contacts were studied and attributed to carrier separation due to band bending at the contacts. Here, however, we investigate photocurrent induced well inside the graphene sample, far from the contacts. While band bending associated with the top gate potential $U_g(x)$ do produce electric fields and hence photocurrents, it is important to understand why the strong photoresponse we observe only appears when a *p-n* junction is formed, not when the gate produces an equally large density gradient in the unipolar (*p-p'* or *n-n'*) regime.

We note that photoexcited carriers typically decay on a time scale of picoseconds, cascading from high energy $hf$ (with $f$ the photon frequency) to a thermal distribution[15]. This occurs well before carriers have reached the contacts, making their direct contribution to photocurrent negligible. However, despite their short lifetime, photoexcited carriers do contribute indirectly to photocurrent response by producing a *local* photocurrent density $j_X$ within the excitation spot, which in turn generates an electric field $E_X = -\rho j_X$, where $\rho$ is the local resistivity. This photoinduced field, $E_X$, then drives current far from the excitation region, which can induce current in the contacts. Because $E_X$ depends on local resistivity, $\rho(n)$, which has a strong peak at $n = 0$, this field is strongly enhanced at the *p-n* junction. In addition, enhanced photoresponse when a *p-n* junction is formed can result from a simple steepening of the potential gradient $\nabla U_g$ due to reduced screening from carriers in the graphene [16].

Besides the conventional photocurrent mechanism, it is necessary to also consider photo-thermoelectric currents, which are also strongly enhanced at *p-n* junctions[17]. In the photo-thermal mechanism, relaxation of photoexcited carriers, through interaction with other carriers and phonons, generate additional excited carriers, yielding a highly populated distribution at a locally elevated effective temperature. Gradients of effective temperature produce thermoelectric fields $E_T = S\nabla T$ by the Seebeck effect ($S$ is the Seebeck coefficient), resulting in photocurrent response at the contacts. Accounting for both mechanisms, we model photocurrent response as the sum of the local photoinduced fields $E_X + E_T$, integrated over the sample area,

$$I = R_{SD}^{-1} W^{-1} \iint \left( -\rho(n)\eta N_X \partial_y U_g + S\partial_y T \right) dx. \quad (1)$$

In Eq. (1), $R_{SD}$ is the source-drain resistance, $W$ is the width of the top gate along the p-n interface, and $N_X$ and $\eta$ are density and mobility of the photoexcited carriers[17]. According to the Mott formula, $S$ is proportional $d\ln(\rho)/d\mu$, where $\mu$ is the chemical potential. Thermopower measurements confirm that this relation holds in graphene[11]. Importantly, $S$ depends on density and strongly varies near zero density. Although the temperature gradient has zero mean when taken across the whole sample, the strong variation of the Seebeck coefficient in the region where the gradient changes sign can convert the thermal gradients into a sizable net current.

Taking a typical value S=50 $\mu$V/K from Ref. 11, for the measured value of photocurrent $I$=10 nA we estimate the temperature increase at the excitation spot as $\Delta T \sim R \cdot I/S$, arriving at a value $\Delta T \sim 1$ K. Based on a heat balance model, including the 2D thermal conduction in the graphene sheet, this is a reasonable value for the laser power $P$=40 $\mu$W used in our experiment. The photothermal contribution is modeled by assuming a gaussian temperature profile, $T(x) = T_0 + \delta T_0 e^{-(x-x_L)^2/R_L^2}$, where $T_0$=298 K is the external temperature, $x_L$ is the position of the laser spot, $R_L$ is the laser spot size, and the amplitude $\delta T_0 \propto P_0$ is proportional to the laser power $P_0$. In addition, we assume that the electron density changes smoothly over a distance $d$ between the background value $n_0 \sim 2\times 10^{12}$ cm$^2$ and its value $n = C_{tg}(V_g-V_0)$ under the top gate, with gate capacitance $C_{tg}$= 2.5×10$^{16}$ m$^{-2}$ V$^{-1}$, and d ≈ 20 nm, comparable to the distance from the top gate to the graphene sheet. The Dirac point under the top gate occurs at $V_g = V_0 = 0.8$ V; this value controls the point at which photocurrent disappears in the gate voltage direction (see Fig. 2b), and was chosen to be consistent with the position of the Dirac point measured in Fig. 2a. Finally, we model the "rounded V" shape dependence of conductivity on density by $\sigma(n) = \sigma_0 + \sqrt{\delta\sigma^2 + (\eta n)^2}$, with minimum conductivity $\sigma_0 \approx 10^{-4} \Omega^{-1}$, $\delta\sigma = 10^{-5} \Omega^{-1}$, and $\eta = 3.1\times 10^{-21} m^2\Omega^{-1}$ taken from measured device parameters. A more detailed model could include the effect of the substrate and the metal gate on the temperature profile.

For the present device parameters, photothermal currents were estimated to be larger than the photocurrents due to the conventional mechanism. Separating these contributions experimentally has not been done, but could perhaps be accomplished using a pulsed laser, as the direct mechanism leads to a faster response than the thermal one. The characteristic crossover pulse duration, below which the conventional mechanism dominates, depends on the details of the graphene-substrate interface, and has not yet been estimated or observed. Photocurrents calculated using Eq. (1) are shown in Fig. 2c and Fig.

3 using experimentally determined parameters for the device. Qualitative features as well as the overall magnitude of the effect are robust over a range of parameters and are in good agreement with experiment.

When the density outside the top-gate region is tuned near the charge neutrality point, denoted *0,* the on-off response to creating a *p-n* interface is replaced with a symmetric response around charge neutrality under the gate, as shown in Fig. 3. Sweeping the top gate across the Dirac point changes the overall device configuration symmetrically, from *0-p-0* to *0-n-0*. In this case, the photocurrents generated along top-gate edges reverses across this transition.

The largest observed gate-dependent photocurrent was found in a bilayer device, shown in Fig. 4, for $\lambda$ = 532 nm and P = 30 μW. Spatial dependence of the photocurrent for the bilayer device in the *n-n'-n* regime ($V_T$ = -1 V), with weak (nearly absent) photoresponse at the gate edges, and in the *n-p-n* regime ($V_T$ = -10 V), where the photoresponse at the gate edge is strong, is shown in the insets of Fig. 4. Responsivity, defined as drain current per watt of incoming radiation, measured at two points on the device, at the edge of the top gate and the edge of one contact, are shown in the main panel of Fig. 4. At the top gate edge, responsivity depends strongly on top gate voltage, turning on when a *p-n* junction is formed, and reaching 1.5 mA/W at $V_T$ = -10 V. This is comparable to the previous record for unbiased graphene photodetection[18]. The high responsivity with top gated devices reported here is achieved at zero bias and thus any dark current is absent. These results confirm that the underlying photothermal physics are similar in bilayer and single layer graphene devices. In addition, higher photocurrents can be achieved in bilayer graphene due to the higher light absorption by a factor of two.

In conclusion, by tuning both single-layer and bilayer gated graphene devices from bipolar to unipolar, we have demonstrated gate-activated photoresponse. Our results are consistent with a model of the photothermal effect, where elevated temperature at the *pn*-junction induces thermoelectric current through the junction. We anticipate that the responsivity can be further increased by converting incoming light more efficiently to a thermal gradient by integration with metallic plasmonic structures, or by reducing the device size, using transparent top gates, and by optimizing device technology to enable p-n devices in the ballistic regime[19]. Neither of the photocurrent generating mechanisms we have considered, photovoltaic nor photothermal, are limited by a band gap, and so are expected to give broadband gate-controlled response, though this remains to be demonstrated experimentally. With the possible extension into far-IR / THz radiation and the high conductivity of graphene, we envision broadband bolometers with submicron pixelation based on the demonstrated phenomena.

**Acknowledgement**

Research supported in part by NSF/NRI INDEX and ONR through MURI-GATE. M. Lemme acknowledges the support of the Alexander von Humboldt foundation through a Feodor Lynen Research Fellowship, and F.H.L. Koppens acknowledges support from the Fundació Cellex Barcelona.

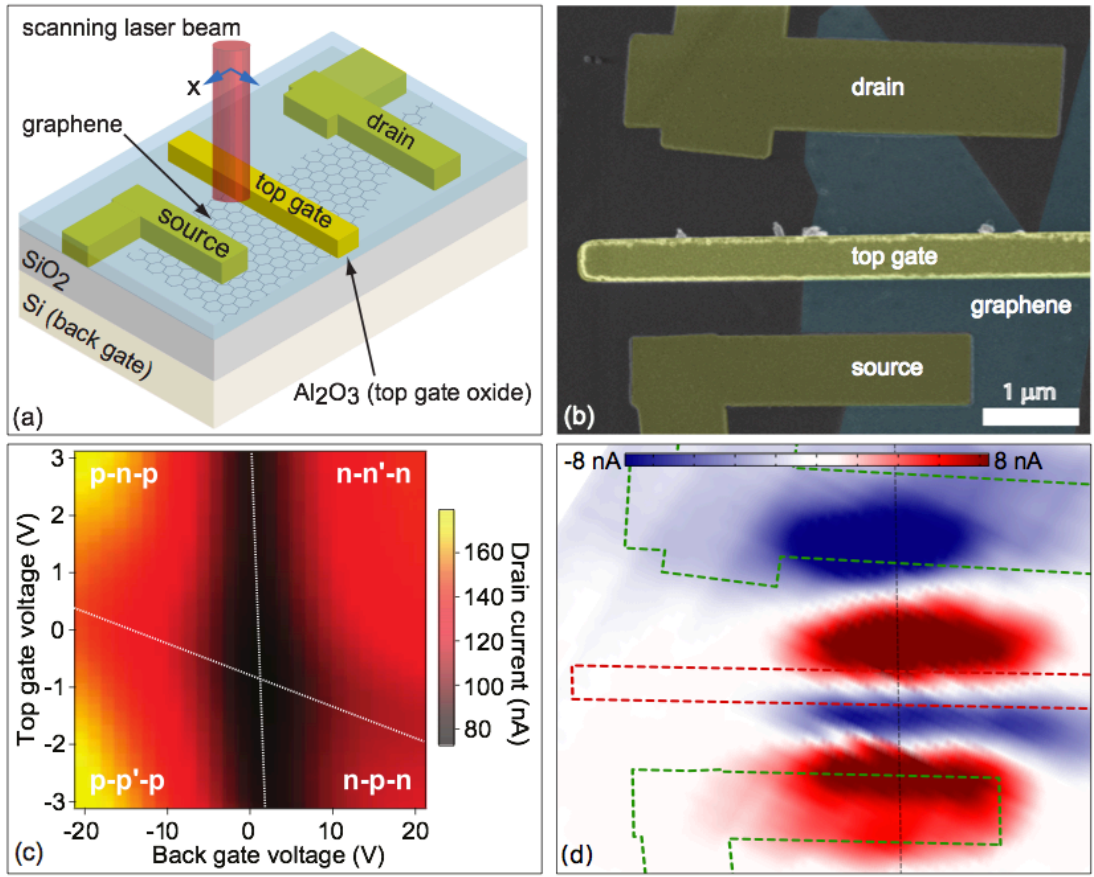

**Fig. 1**: (a) Schematic set-up and (b) false-color scanning electron micrograph of top-gated graphene photodetector. c) Drain current (color scale) as a function of back and top gate voltages. Vertical dashed line indicates charge neutrality point outside of gated region. Diagonal dashed line indicates charge neutrality point under the gated region. d) Scanning photocurrent image of the device in Fig. 1b. Positive and negative photocurrents, measured at the drain contact, appear adjacent to metal leads and the edges of the top-gate.

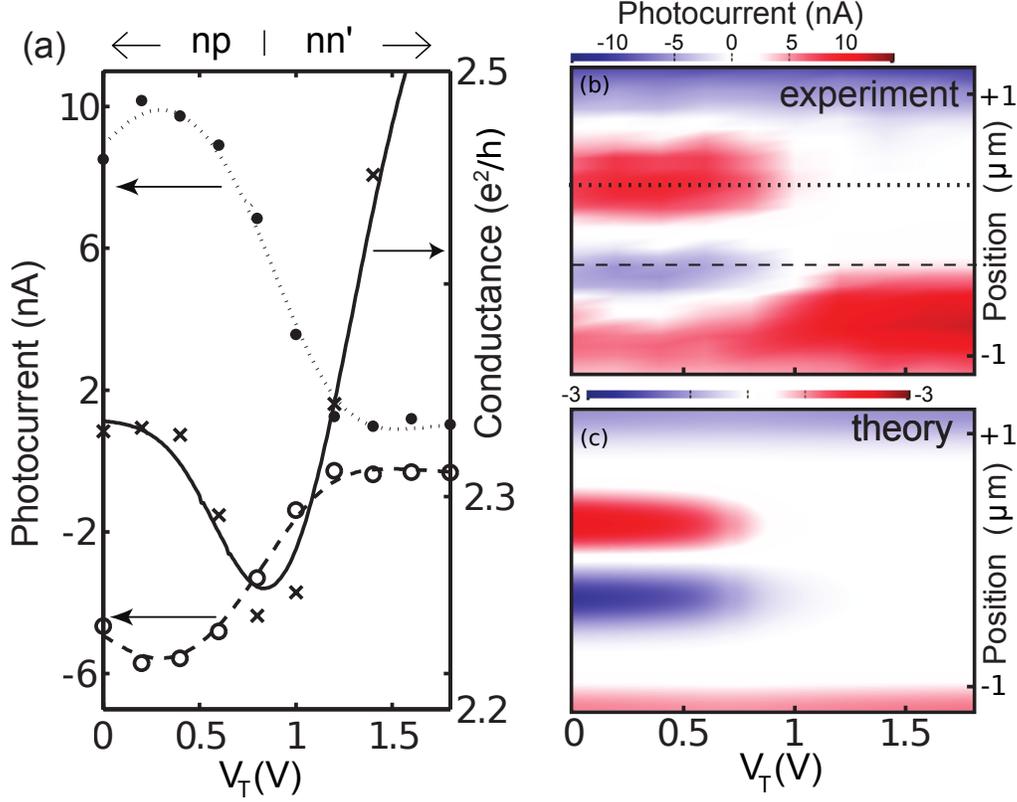

**Fig. 2**: a) Photocurrent (left axis, circle markers) as a function of top gate voltage with the laser positioned on either side of the top gate. Photocurrents turn on at the charge neutrality point under the top gate, as the device is switched from *n-n'-n* to *n-p-n* configuration. Source-drain conductance (right axis, cross markers) of the photodetector measured in FET configuration as a function of top gate voltage with charge neutrality point at $V_T$ = 0.9 V (drain voltage $V_D$ = 0.6 mV). Due to hysteresis when sweeping the top gate voltage, this curve is shifted compared to the data in Fig 1c. b) Photocurrent as a function of top gate voltage taken across the center of the photodetector in Fig 1. The laser wavelength was λ = 600 nm and the power was P = 40 μW. c) Theoretical model of the photocurrent (Eq. 1), plotted as a function of top gate voltage and position along the center of the photodetector. $P_0$ = 40 μW, δT=0.2 K, and we assume 4.6% absorption of the laser light because it passes through the graphene sheet twice due to mirroring at the SiO2/Si interface.

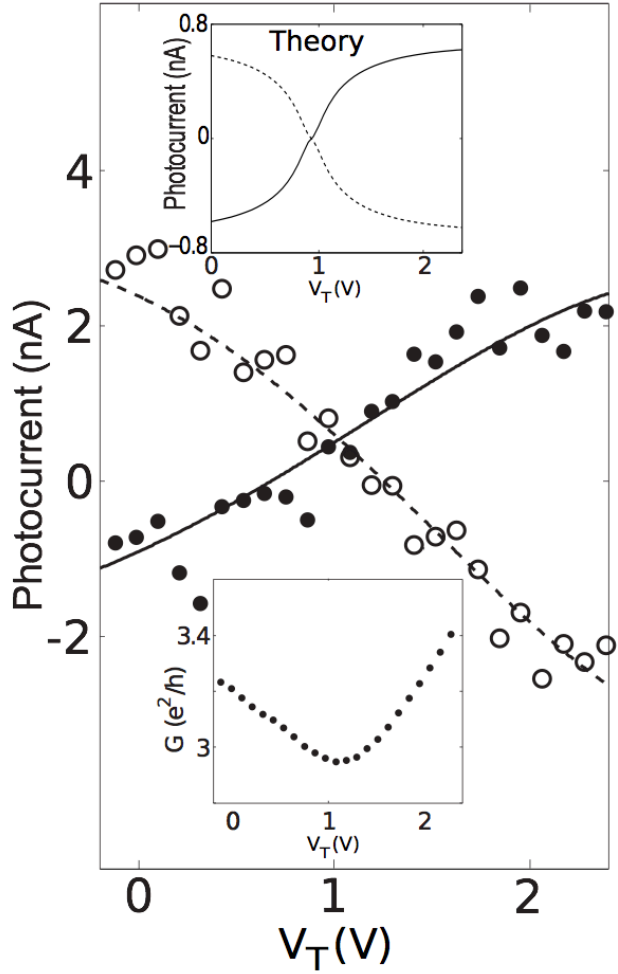

**Fig. 3**: Photocurrent as a function of top gate voltage with the laser ($\lambda$ = 532 nm, P = 62 µW) positioned on either side of the top gate (dots and circles represent one side each). The leads are tuned to the Dirac point by the back gate, resulting in a transition from a *0-n-0* to a *0-p-0* configuration. Bottom inset: corresponding drain current vs. top gate voltage sweep in FET configuration, indicating the location of the Dirac point ($V_D$ = 1 mV). Top inset: theoretical model of the photocurrent, Eq. 1, plotted as a function of top gate voltage, and assuming $P_0$ = 62 µW and zero density away from the top gate.

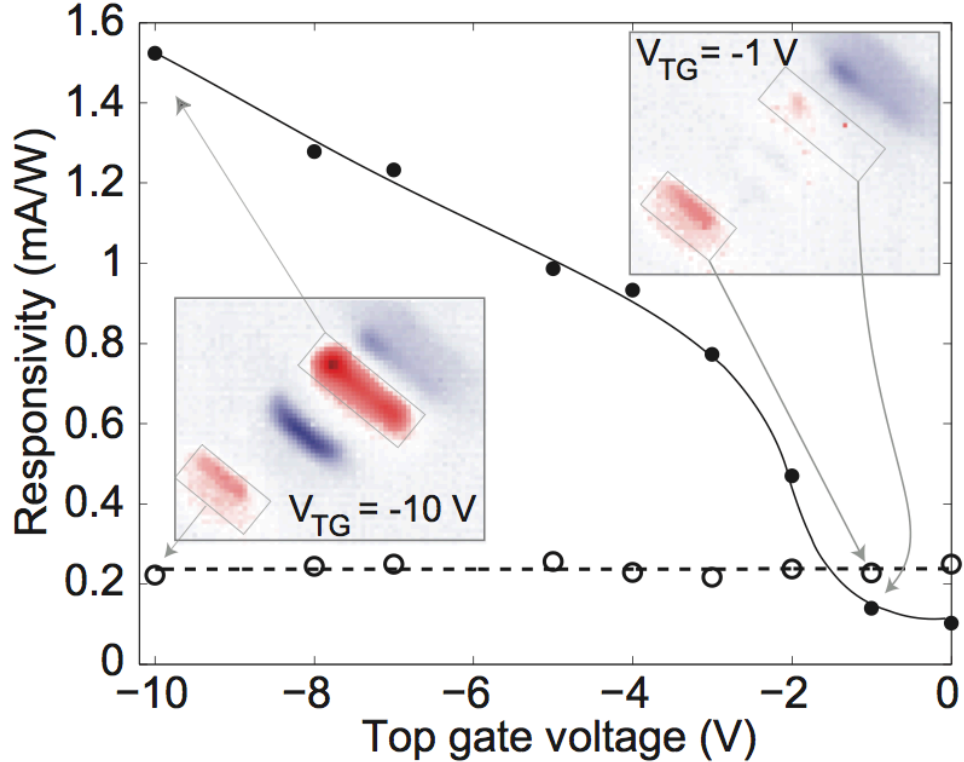

**Fig 4**: Responsivity of gated bilayer photodetector, as a function of top gate voltage with the laser spot ($\lambda$ = 532 nm, P = 30 µW) positioned on the edge of the top gate (solid circles) and on the edge of the metal contact (open circles). Lines are guides to the eye. The sharp increase in responsivity corresponds with a transition from an *n-n'* to an *n-p* junction under the laser ($V_D$ = 10 mV). Insets: scanning photocurrent image at $V_T$ = -1 V (*n-n'-n*) and -10 V (*n-p-n*). The grey boxes indicate the spots where the responsivity was extracted.